\def\bq{\begin{equation}}
\def\eq{\end{equation}}
\def\bqa{\begin{eqnarray}}
\def\eqa{\end{eqnarray}}
\def\bqb{\begin{eqnarray*}}
\def\eqb{\end{eqnarray*}}

\def\mz{M_Z^2}
\def\mw{M_W^2}
\def\sef{s^2_{EFF}(\mz) }

\def\q{q^2}
\def\p{ {\cal P}}
\def\g{\gamma}

\documentstyle[12pt]{article}
\begin{document}
\thispagestyle{empty}

\vspace*{1cm}
\hspace {-0.8cm} PM/93-34 \\
\hspace {-0.8cm} UTS-DFT-93-28\vspace {2cm}
\begin{center}
{\Large\bf Four Fermion Processes at Future $e^+e^-$ Colliders  }
\vspace{0.5cm}\\
{\Large\bf as a Probe of New Resonant Structures  }
    \vspace{1cm}  \\
 {\Large  J. Layssac, F.M. Renard }
  \\
 Physique
Math\'{e}matique et Th\'{e}orique,\\
CNRS-URA 768, Universit\'{e} Montpellier II \\
 F-34095 Montpellier Cedex 5
\vspace {1cm}\\
 {\Large C. Verzegnassi} \\
 Dipartimento di Fisica Teorica,
 Universit\`{a} di Trieste \\
 Strada Costiera, 11 Miramare, I-34014 Trieste, \\
 and INFN, Sezione di Trieste, Italy
\vspace {1.5cm}  \\
 {\bf Abstract}
\end{center}
\noindent
 Possible oblique effects from vector particles that are strongly
coupled to the known gauge bosons are calculated for the case of final
hadronic states produced at future $e^+e^-$ colliders, using a
formalism that was recently proposed and that exploits the information
and the constraints provided by LEP 1 results. Combining the hadronic
channels with the previously analysed leptonic ones we derive improved
limits for the masses of the resonances that, in
technicolour-like cases, would range from one to two TeV
  for a 500 GeV
linear collider, depending on the assumed theoretical constraints.

\setcounter{page}{0}
\newpage

\baselineskip=24pt
 The possibility of using high precision LEP1 data to derive
information, or to set stringent bounds, on technicolour models, has
been thoroughly investigated in recent times, following the original
proposal of Peskin and Takeuchi [1]. As it is known, the relevant
effect is a virtual one-loop contribution, of the so-called [2] oblique
type, to the quantity defined S in ref.[1].\par
 Technically speaking, the calculation of S is made easier by the fact
that the combination of spectral functions that is involved has a
rather exceptional asymptotic convergence, being the difference of a
vector and an axial vector term, and this allows the use of simple
dispersion relations i.e. without unknown extra subtraction constants.
This nice feature would not be present in general in different
kinematical configurations, e.g. away from the Z resonance, for other
oblique corrections of similar type, and an analogous calculation of
technicolour-like effects would require some extra ingredient or ad hoc
assumptions that might bias the theoretical outcome.\par
 In a recent publication [3] we actually proposed a general
 formalism to
calculate the relevant oblique contributions to a number of processes
in future higher energies $e^+e^-$ experiments. The main idea was that
of expressing the various effects in the form of a once-subtracted
dispersion integral, and of fixing the necessary subtraction constants
by suitable model-independent LEP 1 results. In this way, we were led
to a compact "representation" of several observables. In particular, we
concentrated our preliminary analysis on the case of final leptonic
states and more precisely on the three quantities:\par
  a) the cross section for muon production at cm energy $\sqrt{q^2}$,
 $ \sigma_\mu(\q)$.\par
  b) the related forward-backward asymmetry $A_{FB,\mu}(\q)$\par
  c) the (conventionally defined) final $\tau$ polarization asymmetry
$A_\tau(\q)$ or, equivalently, the longitudinal polarization asymmetry
 for final \underline{lepton} production $A_{LR,l}(\q)$ whose
 theoretical
 expressions coincide in our
scheme.\par
 Starting from the tree-level expressions of (a), (b), (c) and making
use of the by now conventional formalism based on the introduction of
the two parameters $\epsilon_{1,3}$, that allows to interpret LEP 1
leptonic data in a model-independent way [4], we were able
 to write for the
oblique (S.E.= self-energy) corrections the following approximate
 formulae, valid
at the one loop level:
  \bqa \sigma_\mu^{SE}(\q)&=&
{4\pi\q\over3}\biggm\{  [{\alpha(\mz)\over q^2}]^2[1+2D_\gamma(\q)]
+ {1\over(q^2-\mz)^2+\mz\Gamma^2_Z}\nonumber\\
 & &\null \bigm[{3\Gamma_l\over M_Z}\bigm]^2[ 1-2 D_Z(\q)
-{16 s^2_1v_1\over1-v^2_1}D_{\gamma Z(q^2)}] \biggm\} \eqa

 \bqa A^{S.E.}_{FB,\mu}(q^2)&=&{3\over4}[{3q^2\sigma_\mu(q^2)\over4\pi}]
              ^{-1}
 \null\biggm\{6\alpha(\mz){\Gamma_l\over M_Z}{q^2(q^2-\mz)\over
 (q^2-\mz)^2+\mz\Gamma_Z^2}\nonumber\\
 & &\null [1+ D_\gamma
(\q)- D_Z(\q)]
\ ]\biggm\} \eqa
\bqa A_\tau^{(S.E.)}(\q)
&\equiv&A_{LR,l}^{(S.E.)}=\bigm[{3\q\sigma_\mu(\q)\over
 4\pi}\bigm]^{-1}A(\mz)
 \null\biggm\{[6\alpha(\mz){\Gamma_l\over M_Z} {\q(\q-\mz)\over
 (\q-\mz)^2+\mz\Gamma_Z^2}\nonumber\\
& &+18({\Gamma_l\over M_Z})^2{q^4\over
  (\q-\mz)^2+\mz\Gamma_Z^2}]
\times[1-{8s^2_1\over A(\mz)} D_{\gamma Z}
(\q)]\ ]\biggm\} \eqa
Here $\Gamma_l$ is the leptonic $Z$ width,
 $\alpha(\mz)=[1\pm0.001]/128.87$[5],
 $A(\mz)$ is defined as
 \bqa A(\mz)\equiv {2(1-4\sef)\over1+(1-4\sef)^2} \eqa
with $s^2_{EFF}(M^2_Z)$ measured by the various asymmetries
at LEP 1 and SLC, and
 \bqa D_\gamma
(\q)\equiv\Delta\alpha(\q)-\Delta\alpha(\mz)
=-{\q-\mz\over\pi}\ \p\int_0^\infty {ds\ Im\ F_\gamma(s)\over
    (s-\q)(s-\mz)} \eqa
  \bqa D_Z(\q)\equiv Re\ [I_Z(\q)-I_Z(\mz)]
 = {\q-\mz\over\pi}\ \p\int_0^\infty {ds\ s\ Im\ F_{ZZ}(s)\over
    (s-\q)(s-\mz)^2} \eqa
  \bqa D_{\gamma Z}
(\q)\equiv Re\ [\Delta\bar \kappa'(\q)-\Delta\bar \kappa'(\mz)]
  = {\q-\mz\over\pi}\ \p\int_0^\infty {ds\ Im\ F_{\kappa'}(s)\over
    (s-\q)(s-\mz)} \eqa
 $$ (F_\kappa'=c_1/s_1\  F_{Z\g}  ,
 s^2_1 c^2_1={\pi\alpha\over\sqrt2 G_\mu\mz}\ \ ,\ \ s^2_1=1-c^2_1\simeq
  0.217 \ \
,\ \ v_1=1-4s^2_1). $$
Eqs.(1),(2),(3) provide a representation of the leptonic observables
of $e^+e^-$ annihilation where the full effect of the oblique
corrections is made explicit in the form of a subtracted
dispersion relation,
thus calculable for models of both perturbative and of non-
perturbative type, with the subtraction constants provided by
model-independent LEP 1
data. Note that, to obtain properly gauge-invariant expressions,
one has still to add the correct amount of extra vertices and
boxes[6], as discussed in ref.[3], to compensate for the intrinsically
not gauge-invariant nature of the transverse self-energies, that are
defined following the convention:
\bq A_{ij}(\q)\equiv A_{ij}(0) +\q F_{ij}(\q)\ \ ,\ i,j=\g ,Z \ . \eq
 Starting from eqs.(1)-(3) and (5)-(7) we calculated in ref.[3] the
possible effects of a couple of vector (V)
and axial vector (A) resonances with masses larger than
$\sqrt{q^2}$, strongly coupled to the photon and to the $Z$. We assumed
a "technicolour-like" framework but only exploited the validity of the
\underline{second} Weinberg sum rule [7]. We did not use the
model-dependent information provided by the first Weinberg
sum rule. However, we
retained one very general consequence of it, i.e.
the positivity of S, which was ensured by the choice $M_A > M_V$.
 Taking into account the LEP
1 constraint [8] on the S-parameter, we derived observability
limits for $M_{V,A}$ in the $TeV$ range for a realistic $e^+e^-$ linear
collider of 500 GeV cm energy [9]. This was an encouraging
preliminary result, particularly since only the final leptonic
channels were fully exploited.\par
 This short paper has two purposes. The first one is that of enlarging
the previous study by including the potentially copious information
provided by the analysis of final hadronic states. The second one is
that of emphasizing the relevance of some special theoretical
assumptions to fix the derived mass limits, in particular of
investigating the consequences of relaxing completely the two
 Weinberg sum
rules, while still retaining the experimental constraint provided
by the LEP 1 limits on the S parameter.\par
 The investigation of the hadronic channels can be easily performed
following the prescriptions of ref.[3]. We shall briefly sketch here
the derivation of the relevant formulae for the "basic" cases of the
two cross sections for production of u-type and d-type quarks,
$\sigma_{u,d}(\q)$.
With this purpose, we start from the expressions of these quantities
at tree level:
$$ \sigma^{(0)}_{u,d}(q^2)=N^{(0)}_{u,d}\ \bigm[{4\over3}\pi\q\bigm]
\bigm\{\bigm({Q_{u,d}\alpha_0\over q^2}\bigm)^2
  +\bigm[{G^0_\mu\sqrt2 M^2_{0Z}\over16\pi}]^2\times $$
 \bq \times
{16[(g_{V,u,d}^0)^2+(g_{A,u,d}^0)][(g_{V,l}^0)^2+(g_{A,l}^0)]\over
 D_{0Z}^2 }]
 -2Q_{u,d}{\alpha_0 G_\mu^0\sqrt2 M^2_{0Z}\over16\pi q^2}
4 g_{V,l}^0g_{V,u,d}^0 Re\ {1\over D_{0Z}} \bigm\}  \eq
 where $N_{u,d}$ is the colour factor, $g_{V,A,f}$ are conventionally
defined, i.e. $g_{A_0,f}=T_{3L,f}$ and
$g_{V0,f}=T_{3L,f}-2Q_fs_0^2$,
$G_{\mu 0}$
 is the (bare) Fermi muon decay coupling and $D_{0Z}=\q-M^2_{0Z}$ (the
tree level equality $\alpha_0/s_0^2c^2_0=\sqrt2/\pi\ G_{\mu0} M^2_{0Z}$
 has been used).\par
 When moving to one loop, one has to redefine the Fermi coupling, the
QED coupling, the bare mass $M_Z$, the photon and Z propagators and
the various fermion couplings $g_{V,A,f}$. Then, vertex corrections
and boxes should be correctly included. For the specific purposes of
this paper, that is only dealing with oblique corrections, these terms
will not be explicitely calculated. Thus, in the redefinition of the
Fermi coupling, only the oblique content $A_{WW}(0)/\mw$
will be retained. Analogously,for the
vector couplings we shall
stick to the notations of a previous paper [10] and write, following
essentially the Kennedy and Lynn approach [11]
 \bq {g_{V,f}\over g_{A,f}}=1-4|Q_f|^2s^2_f(\q)
\ \ \ \ \ \mbox{ with }\ \ \
 s^2_f(\q)=s^2_1 [1+\Delta\bar\kappa'_f(\q)] \eq
 The quantity $\Delta\bar\kappa'_f(\q)$
 can be decomposed into a universal self-energy
component $\Delta\bar\kappa'$ and a (light) fermion dependent vertex
correction i.e.(omitting boxes)
\bq \Delta\bar\kappa'_f(\q)=\Delta\bar\kappa'(\q) +\delta'_f \eq
with  $\delta'_f$ ( to be from now on neglected) defined in
ref.[10] and $\Delta\bar\kappa'(\q)$
fixed by the convention
 \bq \sef=s^2_1 [1+\Delta\bar\kappa'(\mz)+\delta'_l ] \eq
 The procedure for deriving compact expressions for the various
self-energy contributions at one loop now follows essentially the
same lines as in the case of ref.[3]. In fact, the pure photon
contribution will generate the usual term $\simeq D_\g(\q)$.
 From the $Z$ contribution,
using the definition[4]
 \bq \Gamma_l={G_\mu M^3_Z\over24\pi\sqrt2}[1+\epsilon_1]
           [1+(1-4\sef )^2] \eq
 \bq \mbox{ with }\ \ \ \
\epsilon_1\equiv -{A_{WW}(0)\over\mw } + {A_{ZZ}(0)\over\mz
}+\mbox{vertices ...} \eq
 an expression containing $\Gamma_l$, $D_Z(\q)$, $D_{\g Z}(\q)$
and $\sef$
 will be originated.
Finally, from the $\g-Z$ interference, a combination of the
previous pure
photon and pure Z case parameters will appear. In practice, the main
difference between the self-energy content of $\sigma_{u,d}$
 and that of $\sigma_\mu$ will
come from the relative weights of the various $D_\g,\ D_Z,\ D_{\g Z}$
 contributions, due to
the various electric charges $Q_f$ that enter both as coefficients of
$\alpha$ and as coefficients of $g_V/g_A$ in eq.(11).\par
 With these premises, it becomes relatively simple to derive the
explicit expressions of the desired one-loop contributions
to $\sigma_{u,d}$.
Neglecting systematically numerically irrelevant contributions
, one obtains the
following simple formulae ( $\sigma^{(1)}_f$ denotes
 the quantity at
one loop):
 \bqb \sigma^{(1)}_{u,d}&=&N^{(1)}_u\ \bigm[{4\over3}\pi q^2\bigm]
\biggm\{ Q_{u,d}^2 [{\alpha(\mz)\over q^2}]^2[
  1+2D_\gamma
(\q)]+\\
 & &\null +{1\over(q^2-\mz)^2+\mz\Gamma^2_Z}
\bigm[{3\Gamma_l\over M_Z}\bigm]^2[
  1+v_{u,d1}^2][1-2D_Z(q^2)]-2Q_{u,d}\alpha(\mz)\times\eqb
\bq \times  {3\Gamma_l\over M_Z}{[q^2-M_Z^2]\over q^2}
{1\over(q^2-\mz)^2+\mz\Gamma^2_Z}v_{u,d,1}v_1[1-({4s_1^2\over v_1}
+4|Q{u,d}|{s_1^2\over v_{u,d1}})D_{\gamma Z}(q^2)]
  \biggm\} \eq
 where $N^{(1)}_f$ is the colour QCD corrected factor and we used the
generalized notation:
 \bq v_{f1}\equiv 1-4|Q_f|s^2_1  \ \ ; \ \ f=u,d \eq
 (and  $ v_{l1}
   \equiv v_1$ ).
 Starting from the "basic" quantities eq.(17) it is now
straightforward to derive the corresponding expressions of a certain
number of hadronic observables. We have considered here the theoretical
expressions
of the following "candidates" to reveal potential self-energy
effects:
\begin{description}
\item[I)]
 $R^{(5)}(\q)$, the ratio $\sigma^{(5)(q^2)}/\sigma_\mu(q^2)$ between
the
 cross sections for production of the
five lighter $(u,d,s,c,b)$ quarks and for muon production
\item[II)]
 $A^{(5)}_{LR}(\q)$, the longitudinal polarization asymmetry for final
 hadronic states of the previous type
\item[III)]
 $R_{b,\mu}(\q)$, the ratio $\sigma_b(q^2)/\sigma_\mu(q^2)$
 between b-quark and muon production
\item[IV)]
$A_{FB,b}(\q)$, the forward-backward asymmetry for $b$ quark production.
 In addition to the previous "old-fashioned" quantities we have also
calculated, assuming a (copious) top production at $\sqrt{q^2} = 500\
GeV$, the theoretical expression of a number of related observables.
In particular, we have considered here:
\item[V)]
 $R^{(6)}(\q),\ A^{(6)}_{LR}(\q),\ R_{t,\mu}(\q)$
 defined in analogy with (I), (II), (III).
\item[VI)]
$R^{(5),(6)}_{b,t}(\q)$, (the ratios
$\sigma_{b,t}(q^2)/\sigma^{(5),(6)}(q^2)$) and $A_{FB,t}(\q)$
\end{description}
 For all the previous observables from (I) to (VI), it is not
difficult to write the expressions at one loop that generalize those
of ref.[3]. But,in the actual process of doing that, one easily
realizes that a priori not all cases seem equally promising. In
particular, assuming "realistic" experimental accuracies (i.e. of the
kind discussed in previous analyses[9]) for the various cross sections
and their ratios, it turns out that the weights of the various
$D_\g,D_Z,D_{\g,Z}$
contributions (that are rather different in the various observables)
are systematically "small" in the cases (IV)-(VI), leading
to practically unobservable effects. For this reason, we
concentrated our attention on the quantities (I)-(III) only. Ignoring as
usually several irrelevant terms, we were led in conclusion to the
following set of expressions that include the full effect of
the oblique corrections at one loop:
\bq R^{(5)(S.E.)}(\q)= a_0[1+a_{\gamma}D_{\gamma}+a_ZD_Z+a_{\gamma
Z}D_{\gamma Z}]                      \eq
\bq R^{(S.E.)}_{b,\mu}(\q)= b_0[1+b_{\gamma}D_{\gamma}+b_ZD_Z
+b_{\gamma
Z}D_{\gamma Z}] \eq
\bq A^{(5)(S.E.)}_{LR}(\q)= c_0[1+c_{\gamma}D_{\gamma}+c_ZD_Z
+c_{\gamma
Z}D_{\gamma Z}]                       \eq
 where the analytic expressions of the various coefficients can be derived in
a straightforward way and their numerical values for $\sqrt{q^2} = 500
 (190) GeV$ are given below:\par
 $a_0=5.59(6.84)$, $a_{\gamma}=-0.61(-0.76)$, $a_Z=-0.84(-1.12)$,
$a_{\gamma Z}=-0.26(-0.32)$ \par
 $b_0=0.88(1.16)$, $b_{\gamma}=-1.10(-1.17)$, $b_Z=-1.21(-1.41)$,
$b_{\gamma Z}=-0.80(-0.78)$ \par
 $c_0=0.61$, $c_{\gamma}=-0.42$, $c_Z=-0.27$, $c_{\gamma Z}=-1.78$ \par
(we only considered the case for $ A^{(5)(S.E.)}_{LR}(\q)$ at a 500 GeV
linear collider)\par
 Starting from the previous expressions eqs.(17)-(19) it is now
straightforward to calculate various kinds of contributions of
self-energy type, in particular that coming from a model that
implies the existence of a couple of strongly coupled vector (V) and
axial vector (A) resonances. For the latter ones we shall follow the
same notations as in ref.[3], adopting the simplest treatment based on
a delta-function approximation (but keeping in mind the discussion
given there on the possibility of using a more realistic description
without changing the essential results i.e. the mass limits). We
shall not abandon at this stage the customary assumption of isospin
and parity conservation. Thus, the imaginary parts of the various
spectral functions will simply be expressed in terms of the two
 quantities
$R_{VV}, R_{AA}$ with
\bq R_{VV,AA}= 12\pi^2 F_{V,A}^2 \delta (s-M_{V,A})   \eq
Our investigation now proceeds in two steps. First, we assumed as we
did in Ref.[3] the validity of the two Weinberg sum
rules (but only fully exploited the consequences of the second one)
 and  we made use of the experimental constraint on the
parameter $S$ ,that can be written to quite reasonable an approximation
as:
\bq -1.5 \leq \ S\ \leq 0.5 \eq
only considering the \underline{positive} upper bound. Then we
 combined the previous ansatzs with the request that the
experimental accuracies on $R^{(5)},A^{(5)}_{LR}$ and $R_{b,\mu}$ are
of a relative one,one and two percent respectively [9] and imposed
the consequent
"observability" limits.\par
Fig(1) shows the results of our analysis for the case
 $\sqrt{q^2}=500 \ GeV$.
The different curves correspond to the various observables, and
the shaded area corresponds to the combined overall mass bound.
\par
{}From inspection of Fig.(1),the following main conclusions may be
derived:
\par
a) the only hadronic observable which contributes
 appreciably the
bound is $A^h_{LR}$, that allows to improve the pure leptonic result
by approximately 150 GeV.    \par
b) the resulting bounds on $M_V,M_A$ are located in the TeV range,
and rather strongly correlated. For the QCD-like choice $M_A/M_V =
1.6$, values of $M_V$ up to 1 GeV would be seen.\par
In the previous analysis, several theoretical assumptions (or
prejudices ?) were inforced, on which the obtained bounds certainly
 depend. To try to make the interconnection between
the numerical output and the theoretical input more quantitatively
defined might be an interesting goal. With this aim, we considered the
consequences of abandoning some of the starting ingredients of our
approach. Since we would personally feel uneasy in giving up the
familiar isospin and parity conservation philosophy, we began by rather
eliminating the assumptions of validity of
both Weinberg sum rules and only retained a "minimal" convergence
assumption $(F_{VV}(q^2)-F_{AA}(q^2)) \sim 0 , q^2 \to
\infty $, to ensure the unsubtracted form of S.
 This choice has two main consequences,
that of introducing another degree of freedom in the analysis and that
of allowing the Peskin-Takeuchi parameter $S$ to become negative,
since one has now
\bq S= \biggm[{F^2_V\over M^2_V}-{F^2_A\over M^2_A}\biggm] \eq
with no special indications for its sign. Thus, the experimental
constraint for $S$, eq.(42),
will now allow both end points of the allowed interval to be
saturated.\par
In performing our numerical analysis, we had to solve the problem of
the presence of one additional degree of freedom. We decided to
proceed by retaining
a "prejudice relic" in which the value of the ratio $F_V/M_V$ was
bounded by the limit
\bq {F_V\over M_V}=2{f_\rho\over m_\rho}={1\over\sqrt{2\pi}} \eq
i.e. twice the QCD value. Higher values of the ratio would obviously
increase the mass bounds accordingly, as from eq.(31). Then ,
 for every choice of $F^2_V/M^2_V$,
$F^2_A/M^2_A$ was allowed to saturate both limits of eqs.(42). The
final results were then plotted as in the case of Fig.(1) in the
$(M_V,M_A)$ plane. In Fig.(2) we give the results of the procedure that
correspond to the choice $F^2_V/M^2_V=1/2\pi$, showing that the
 situation
has now definitely changed with respect to Fig.(1). In particular, one
sees now that
 the effect of releasing the validity
of the Weinberg sum rules is roughly that of increasing the bounds on
$(M_V,M_A)$ from
the 1 TeV region to the 2 TeV region for a reasonable limitation
 on $F_V/M_V$. The effect of the hadronic observables is still to
increase the mass bounds by about 150 GeV.\par
To complete our analysis, we examined the similar sitation
 that would
occur at $\sqrt{q^2}=190\  GeV$, i.e. the near future LEP2 energy.
 We proceeded
as before with the experimental conditions expected by previous
analyses [12]. The results that we obtained are shown in Fig.(3)
. As one sees LEP2 under realistic
experimental conditions would be able to reveal signals of strong
resonances whose masses range up to 300-350 GeV (assuming the Weinberg
sum rules) or to 400-450 GeV (releasing them). These values appear
relatively low in classical TC pictures [13],but would certainly be
much more interesting in non orthodox TC versions more recently
suggested [14] implying the existence of 'light"
strongly resonant states.\par
 In conclusion, and although our investigation
 was relatively qualitative, we feel that its
indications should be considered as an example of the
potential interest of such measurements at future $e^+e^-$ colliders.
\newpage

\def\pr#1#2#3{ Phys. Rev. ${\bf{#1}}$, #3 (#2) }
\def\prl#1#2#3{ Phys. Rev. Lett. ${\bf{#1}}$, #3 (#2) }
\def\pl#1#2#3{ Phys. Lett. ${\bf{#1}}$, #3 (#2) }
\def\sjnp#1#2#3{ Sov. J. Nucl. Phys. ${\bf{#1}}$, #3 (#2) }
\def\np#1#2#3{ Nucl. Phys. ${\bf{#1}}$, #3 (#2)  }
\def\zp#1#2#3{ Z. Phys. ${\bf{#1}}$, #3 (#2) }

\centerline{{\bf References }}
\begin{description}

\item[{[1]}] M.E. Peskin and T. Takeuchi, \pr{D46}{1991}{381}.

\item[{[2]}] B.W. Lynn, M.E. Peskin and R. Stuart in "Physics at
LEP", J. Ellis and R. Peccei eds., CERN 86-02 (1986), Vol 1.

\item[{[3]}] J. Layssac, F.M. Renard and C. Verzegnassi,
\pr{D48}{1993}{4037}.

\item [{[4]}] G. Altarelli, R. Barbieri and S. Jadach,
\np{B369}{1992}{3}.

\item [{[5]}] H. Burkhard, F. Jegerlehner, G. Penso and C. Verzegnassi,
\zp{C43}{1988}{497}

\item[{[6]}] G. Degrassi and A. Sirlin,\pr{D46}{1992}{3104}, for the
original proposal see also J.M. Cornwall,\pr{D26}{1982}{1453}.

\item[{[7]}] S. Weinberg,\prl{18}{1967}{507}.

\item[{[8]}]  See e.g. the talk given by G.Altarelli at the Marseilles
Conference, July 1993.

\item[{[9]}] See e.g.: $e^+e^-$ collisions at 500 GeV: the Physics
Potential", DESY 92-123-1992, edited by P.M. Zerwas.

\item[{[10]}] J. Layssac, F.M. Renard and C. Verzegnassi, preprint
UCLA/93/TEP/16, to
appear in Phys.Rev. D.

\item[{[11]}] D.C. Kennedy and B.W. Lynn, \np{B322}{1989}{1}.

\item[{[12]}] A. Blondel, F.M. Renard, P. Taxil and C. Verzegnassi,
\np{B331}{1990}{293}.

\item[{[13]}] S. Weinberg, \pr{D13}{1976}{974}, {\bf D19} (1979)
1277;
 L. Susskind, \pr{D20}{1979}{2619};
 E. Fahri and L. Susskind, \pr{D20}{1979}{3404}.

\item[{[14]}] R.S. Chivukula, M.J. Dugan and M. Golden,
\pl{B292}{1992}{435}.

\end{description}
\newpage

\par
\centerline { {\bf Figure Captions }}\par
 Fig.1 Limits on $M_A$ at variable $M_V$
 obtained at $\sqrt{\q}=500\ GeV$
from  $\sigma_\mu$ (dotted), $A_{LR,h}$ (dot-dashed)
 and $A_{\tau}$ (dashed), using the Weinberg sum rules and the
experimental information on S.
 The lighter shaded domain represents the result of combining
quadratically the two leptonic limits. The darker one corresponds to
the domain allowed by the leptonic and the hadronic limits. The two
full lines correspond to $M_A=1.6 M_V$ and to $M_A=1.1 M_V$. \par
 Fig.2 Limits when releasing the Weinberg sum rules but
imposing the limitation on $F_V/M_V$, from $\sigma_\mu$
(vertical,dotted), $A_{\tau}$ (vertical, dashed),
  $A_{LR,h}$ (dot-dashed), $R_{b,\mu}$ (short dashed), $R^{(5)}$
(dotted), $A_{FB,\mu}$ (long dashed). The shaded domains have the
same meaning as in Fig.1. The two full lines now correspond to
$M_A=1.6 M_V$ and to $M_A=M_V$. \par
 Fig.3 Resulting domains obtained at $\sqrt{\q}=190\ GeV$ (same
meaning as in Fig.2)
with accuracies expected at LEP2.

\end{document}